\documentclass[%
 aip,
 amsmath,amssymb,
 reprint,%
]{revtex4-2}

\usepackage{graphicx}
\usepackage{dcolumn}
\usepackage{bm}

\usepackage[utf8]{inputenc}
\usepackage[T1]{fontenc}
\usepackage{mathptmx}

\begin{document}

\preprint{APL}

\title {A perspective on integrated atomo-photonic waveguide circuits}

\author{Yuri B. Ovchinnikov}
\affiliation {National Physical Laboratory, Hampton Road, Teddington TW11 0LW, United Kingdom}

\date{\today}

\begin{abstract}
Integrated photonic circuits based on suspended photonic rib waveguides, which can be used for coherent trapping, guiding, and splitting of ultra-cold neutral atoms in two-colour evanescent light fields near their surfaces, are described. Configurations of quantum inertial sensors based on such integrated atomo-photonic waveguides, which are simultaneously guiding photons and atoms along the same paths, are presented. The difference between free-space and guided atom interferometers in the presence of external forces is explained. The theoretical and technological challenges, to be overcome on the way to the realization of such a novel platform for quantum technologies, are discussed.
\end{abstract}

\maketitle

Just as the invention of semiconductor lasers in the 1960s led to the development of photonic circuits, the development of Bose-Einstein condensate (BEC) of dilute atomic gases  \cite{1} and corresponding atom lasers  \cite{2,3} in the 1990s is now leading to the development of integrated circuits for ultra-cold atoms. Although the quasi-monomode guided atom laser already has been established \cite{4}, its integrated version has still to be developed. As with photonic and electronic circuits, all guiding circuits and their elements for ultra-cold atoms are unified now under the new term of atomtronics \cite{5}. The main advantage of atomic circuits is that they are combining certain properties of both photonic and electronic circuits. On the one hand, taking into account that atoms have rich internal energy structure and large mass, compared to electrons and photons, atomtronics can provide new possibilities for corresponding elements of quantum logic and quantum sensors. On the other hand, the cold-atom-based elements similar to electronic components such as diodes, transistors, and SQUIDs can be developed. Taking into account that atoms, compared to photons and electrons, have a richer structure of internal energy states, we can expect the future invention of principally new quantum components of atomtronics. Note that many atomtronic circuits are not integrated, but rather based on specially designed optical dipole potentials in free space. One example of it are painted optical dipole potentials formed by rapidly moving along certain trajectories focussed laser beams \cite{6}.

Currently existing integrated platforms for ultra-cold atoms, so called "atom chips" \cite{7,8}, are mainly based on magnetic trapping or guiding of atoms above integrated current-caring wires or permanent magnetic structures.

All-optical atom chips, which are now under active development, based on using conservative optical dipole forces for different coherent manipulations with confined atoms, have several advantages over magnetic atom chips. The sizes of individual elements of such chips, like traps, waveguides, and beam splitters, can be comparable to the wavelength of light, leading to overall miniaturization and the provision of additional opportunity for interaction between atoms trapped in adjacent individual traps. The optical dipole traps can confine atoms in magnetically insensitive internal states, which makes these chips attractive to applications, in which the influence of environmental magnetic fields is unwanted. Using light provides easier, more varied, and faster manipulations with atoms. Finally, in absence of any essential absorption of light in photonic structures of an all-optical atom chip, there is no need for additional cooling.

There are at least two different approaches to implementation of all-optical atom chips. The first approach is based on the conservative trapping and guiding of atoms in non-resonant two evanescent light fields of different colour and penetration depth \cite{9}.

There were several proposals and attempts to create surface waveguides for atoms \cite{10,11,12,13,14} based on two-colour evanescent light waves generated above dielectric ridge optical waveguides of subwavelength transverse sizes. One problem with planar waveguides on top of a dielectric substrate is the small penetration depth of evanescent light waves in a vacuum, which makes it possible to form the corresponding evanescent light traps at reasonably large distances from the dielectric surface only at very high intensities of the propagating light. Another problem consists in a rather poor lateral confinement of atoms in evanescent light waveguides of subwavelength transverse size \cite{15}.   

The problems mentioned can be efficiently solved with suspended optical rib waveguides \cite{16,17,18}, which are being actively developed for different kinds of optical sensors based on evanescent light fields. The main advantage is large penetration depths of evanescent light waves in the surrounding medium of suspended waveguides, which can be explained by the equally high contrast of the refractive index from both sides of a suspended planar optical waveguide. This approach is also preferable for trapping of atoms in two-colour evanescent light fields, as it moves the bottom of corresponding optical dipole potentials further away from the dielectric surface of the photonic waveguide. This in turn reduces necessary powers of optical fields for stable trapping of atoms \cite{15} in presence of their Casimir-Polder attraction to the dielectic surface of the waveguide \cite{19}. Modelling of trapping potentials near the surface of suspended optical rib waveguides, which includes optical dipole potentials of two-colour evanescent light waves and the attractive Casimir-Polder potential \cite{15}, shows that it should be possible to realise stable conservative trapping and guiding of atoms in such potentials with a coherence time of guided or trapped atoms up to one second.

Recently a suspended optical rib waveguide based on aluminium oxide membrane, which was intentionally designed for atomic trapping applications, was demonstrated and tested \cite{20}. It is remarkable, that this suspended waveguide has rather low losses in visible and NIR optical spectrum and can withstand powers of the guiding light up to 30 mW, which is more than enough for corresponding optical trapping and guiding of atoms. Another advantage of this approach is that it has a very good optical access to the waveguide due to the large total area of the membrane, which is supposed to provide convenient loading of the waveguide with ultra-cold atoms from any external atomic trap, like magneto-optical traps (MOT), optical dipole traps or magnetic traps.

Another promising technology with several advantages for suspended optical waveguides is based on silicon dioxide membranes on top of silicon substrates \cite{18}, applications of which in all-optical atom chips were modelled earlier \cite{15}. First, silicon dioxide is the main material for most optical fibres due to its very high transparency for visible and NIR light. Second, the purity of a silicon oxide membrane produced by thermal oxidation of a silicon wafer can be very high. Third, a silicon wafer, being a crystal, can have both very high flatness and very low roughness. As a result, it can provide the same high quality of the oxidized layer, silicon dioxide, on a surface of the silicon wafer. Fourth, all the technologies of silicon wafers processing are well established.

The design of a suspended silicon dioxide rib waveguide on top of a silicon wafer is shown in Fig. 1. The two parallel rows of rectangular openings in the SiO$_2$ layer from both sides of the rib waveguide are used for etching silicon from the bottom of the membrane \cite{18}. This process provides comparable transverse sizes of the membranes and depth of wells under them, which provides their mechanical stability. This method can be used for production of crossed and curved suspended waveguides, which is important in their application for integrated waveguide atom interferometers.

\begin{figure}
	\centering
	\includegraphics[scale=0.5]{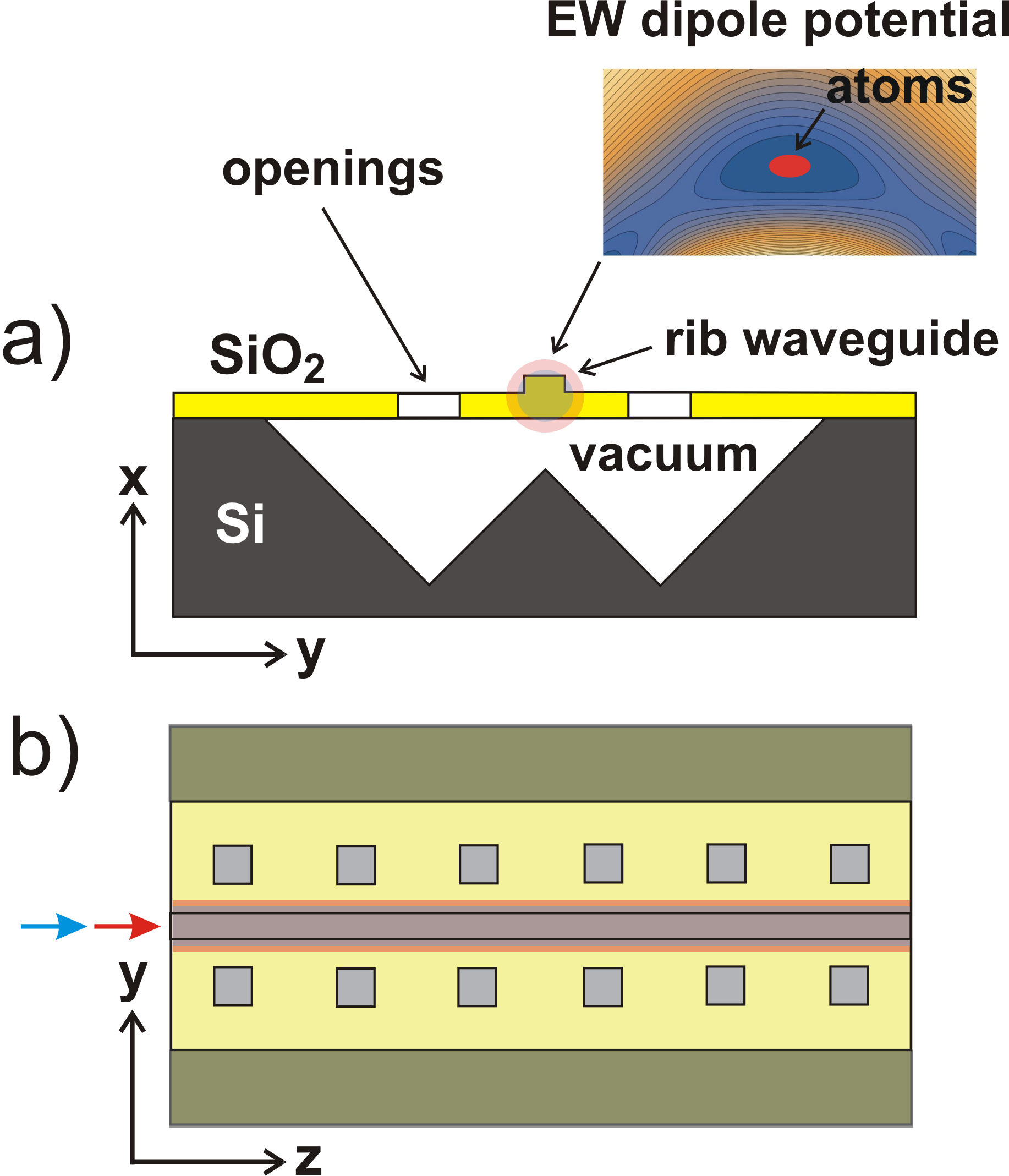}
	\label{}
	\caption{Design of suspended optical rib waveguide for guiding of atoms in two-colour evanescent light waves. a) Cross-section of the waveguide. Inset shows a magnified optical dipole potential near the surface of the rib. b) Top view of the suspended rib waveguide.}
\end{figure}

Geometric parameters of the suspended rib waveguide should support a single-mode regime of propagation of two-colour laser light, the evanescent waves of which are used to produce guiding optical dipole potentials for atoms \cite{15}. Minimums of these potentials are typically located at a distance of several hundred nanometres from a surface of corresponding photonic waveguides. Therefore, the guided ultra-cold atoms are propagating along the integrated optical waveguides. Photonic waveguides which simultaneously guide both photons and atoms along parallel trajectories may be called atomo-photonic waveguides (Fig. 2). These atomo-photonic waveguides can be curved, which opens new possibilities for the design of corresponding atomtronics circuits and atom interferometers.

\begin{figure}[h] 
	\centering
	\includegraphics[scale=0.5]{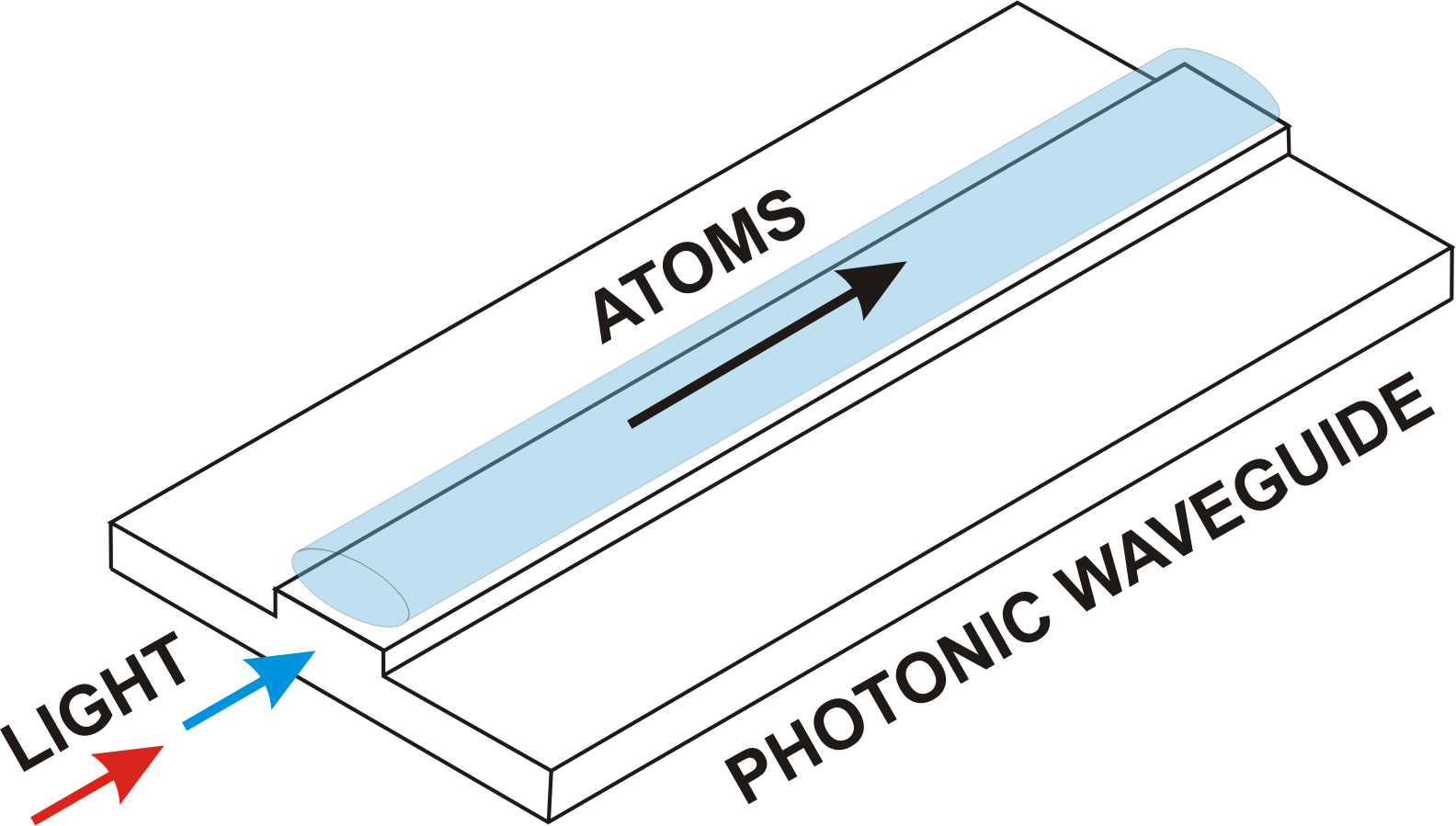}
	\caption{Principle of atomo-photonic waveguide based on guiding of atoms in two-colour evanescent light fields along the surface of a photonic waveguide.}
	\label{fig_rib_section}
\end{figure} 

An important element of the atomic waveguide interferometer is a beam splitter, which splits an initial matter-wave coherently between two or more crossing waveguides. The most common atomic beam splitters are based on coherent diffraction of atoms in laser fields \cite{19}, where the initial atomic wave is split in several partial atomic waves with different transverse momentums. Waveguide all-optical atom interferometers are based on quasi-Bragg waveguide beam splitters, which are formed by optical lattices produced by interference between laser fields of crossed photonic waveguides \cite{15}. This special regime of diffraction corresponds to a Bragg diffraction of high order at an optical lattice of relatively large amplitude and spatial width \cite{20,21}, which makes it different from a classical Bragg diffraction at a shallow and thin optical grating \cite{22}. To date the splitting of atoms in all-optical waveguide quasi-Bragg beam splitters was studied for intersected multimode waveguides based on Gaussian laser beams, where the process of splitting usually led to excitation of higher transverse modes of atomic waveguides \cite{20}. In single-mode atomo-photonic waveguides \cite{15}, such a splitting of atoms between two crossed waveguides cannot lead to the excitation of higher transverse modes of the waveguides. However, it can lead to some loss of guided atoms, which is still a subject for additional theoretical studies.

We now discuss the possible designs and performances of waveguide atom interferometers based on intersecting atomo-photonic waveguides shown in Fig. 3. 
To estimate sensitivities of such waveguide interferometers to external forces we use Feynman's theory of path integrals \cite{23}. According to this theory, the accumulated phase of the atomic wave function over the given pass is proportional to the classical action, which is equal to the integral of Lagrangian over the given path. In waveguide atom interferometers the paths of least action are lying near the bottoms of the guiding potentials. Therefore, the corresponding accumulated phase shift of a matter wave along each of the arms of the interferometer due to an external potential is given by \cite{22}

\begin{equation} \label{EQ1} 
\phi_{i}=\int_{\Gamma_i}{\left[\sqrt{\frac{2m}{\hbar^2}\left(E-U(\vec{r})-U_{wg}\right)}-\sqrt{\frac{2m}{\hbar^2}\left(E-U_{wg}\right)}\right]}dl,
\end{equation}
where $i=1,2$ is an index corresponding to the two arms of the interferometer, $E$ is the total energy of an atom, $U(\vec{r})$ is a potential of the external field and $U_{wg}$ is a value of a guiding optical dipole potential near the bottom of the atomo-photonic waveguides and $m$ is the mass of the atom. For simplicity, we suppose that the potential depth of the optical waveguides $U_{wg}$ is the same and constant along both arms of the interferometers. A second term under the integral in equation (1) corresponds to a wave vector of an atomic matter wave inside the waveguide in absence of external potentials $k_{wg}=\sqrt{\frac{2m}{\hbar^2}\left(E-U_{wg}\right)}$, which is related to the velocity of propagation of atoms inside the waveguide as $v_{wg}=k_{wg}\hbar/m$. In our simplified consideration we neglect the phase shifts of the two partial atomic matter waves at the two Bragg beam splitters of the interferometer. Note, that compared to atomic interferometers based on free-falling atoms, in the waveguide optical interferometers atoms do not change their position relative to the optical lattice, which provides their momentum splitting.
In a case when the optical standing waves at both beam splitters of the waveguide Mach-Zehnder interferometer (Fig. 3a) have the same phases, they should not influence the considered above integrated phase difference between the two passes of the interferometer.

\begin{figure}[h] 
	\centering
	\includegraphics[scale=0.5]{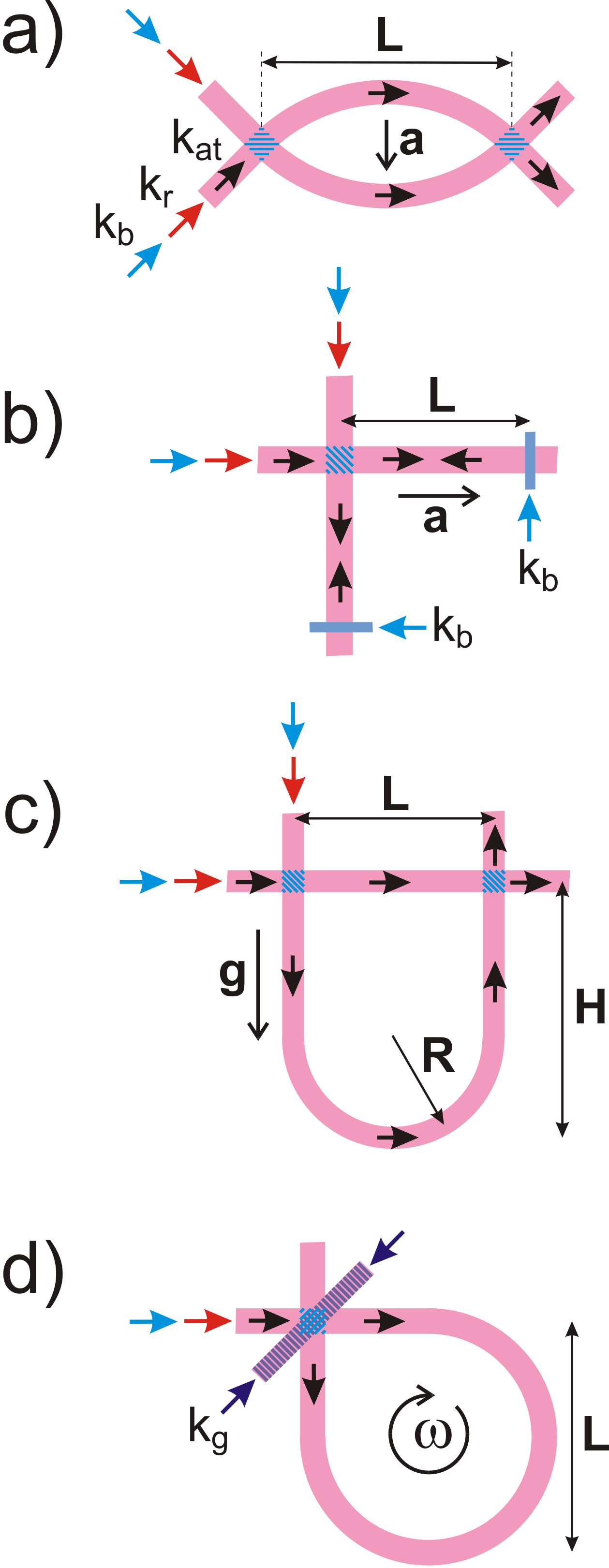}
	\caption{Main configurations of atom waveguide interferometers based on atomo-photonic waveguides. a) Mach-Zehnder interferometer; b) Michelson interferometer; c) asymmetric Mach-Zehnder interferometer for measurement of gravity; d) Sagnac interferometer.}
	\label{fig_rib_section}
\end{figure}

Fig. 3a shows a Mach-Zehnder waveguide interferometer, the enclosed area of which is formed by two intersected curved waveguides, both of which are arcs of a circle with radius $R$. The waveguides intersect each other at an angle of $\pi/2$ at both ends of the interferometer. It means that the central angles of the two arcs are also equal to $\pi/2$ and the length of each of them is equal to $\pi R/2$. The two waveguide Bragg atomic beam splitters of the interferometer are produced by optical lattices formed by interference of evanescent light waves in the cross regions of the interferometer \cite{15,20}. A small transverse acceleration $a$ of the interferometer, according to the formula (1), leads to an additional difference in the accumulated phase $\Delta\phi=\phi_1-\phi_2=\sqrt{2}maL^2/\hbar/v_{wg}$ between the two arms of the interferometer. Based on that phase shift it is possible to estimate the quantum projected noise (QPN) limited sensitivity of the interferometer to acceleration. We assume also that the initial atomic wave packet is split equally between the two arms of the interferometer and the visibility of the interference fringes is equal to 100$\%$. For the length of interferometer $L=1$ cm, the radius of curvature of the arms $R=L/\sqrt{2}=0.707$ cm, initial velocity of atoms in the waveguides $v_{vw}=1$ cm/s and the flux of atoms $10^5$ at/s, the corresponding sensitivity is $S_a=2.6\times10^{-10}$ m/s$^2$/$\sqrt{\text{Hz}}$. Such a high sensitivity of the interferometer is achieved due to the large separation of the arms of the interferometer and the long interrogation time of $T=1.11$ s. Due to the Sagnac effect, such a Mach-Zehnder interferometer is also sensitive to its rotation in the plane of the interferometer. The rotation of the interferometer around its centre with an angular velocity $\omega$ leads to the accumulated phase difference between atoms in two arms of the interferometer of $\Delta\phi=0.5(\pi-2)\omega m L^2/\hbar$. The corresponding sensitivity to rotation is $S_\omega=6.35\times10^{-8}$ rad/s/$\sqrt{\text{Hz}}$.      

A Michelson waveguide atom interferometer is shown in Fig. 3b. The interferometer is composed of two straight waveguides crossed with each other at right angles. The cross-region forms an optical lattice, which is used as a beam splitter for atoms. Each of the waveguides ends with sharp transverse repulsive optical dipole potentials, which are used to reflect the propagating atoms towards the beam splitter. In practice, these reflective potentials can be realised with additional transverse waveguides contain only a single frequency light mode, which is detuned to the blue side of the main atomic transition. Such a waveguide interferometer will measure the difference between accelerations along its two arms. Note, it is not sensitive to its rotation around the centre of its beam splitter if the accelerations are small and both arms have the same length. Let us consider that some small acceleration $a$ is directed just along one of the interferometer arms. The corresponding phase difference between the two partial atomic matter waves propagating along the two arms in presence of that acceleration is $\Delta\phi=maL^2/\hbar/v_{wg}$, where $L$ is the length of arms.  For $L=0.5$ cm and the flux of atoms $F=10^5$ at/s, the sensitivity of such an interferometer to acceleration is $S_a=1.45\times10^{-9}$ m/s$^2$/$\sqrt{\text{Hz}}$.

The above two examples of the waveguide interferometers assumed that the external potential is very small compared to the total energy of the atom. A situation when the external potential is comparable or bigger than the total energy of atoms is very different. Compared to interferometers based on free-falling atoms, large external forces in the waveguide interferometers lead to different accelerations of atoms in the two arms of the interferometers. This leads to a difference in propagation times of atoms along the two interferometer arms. Interference between the atoms after their return to the beam splitter can happen only if the difference between the propagation times of atoms along two arms satisfies the condition $\Delta t v_{wg} <l_c$, where $l_c$ is the coherence length of the atomic wave packet \cite{24}. Therefore, to provide interference of atoms in the case of large accelerations, the length of waveguide interferometer arms should be different. For example, in the case of the Michelson waveguide interferometer (Fog. 3b), the arm length $L_2$, along which the acceleration is applied, must be longer than the length of the other arm by $L_2-L_1=0.5 a L_1^2/v_{wg}^2$. In principle, such a Michelson interferometer can be used as a gravimeter, if its longer arm is placed vertically and directed downwards.

A Mach-Zehnder waveguide interferometer also can be used as a gravimeter. An example of such a gravimeter is shown in Fig. 3c. This interferometer consists of two intersecting atomo-photonic waveguides, while one of them is a horizontal straight waveguide of internal length $L$ and a U-shaped bottom waveguide of total height $H$. Such a configuration of the interferometer provides the same velocity of atoms $v_{wg}$ at both waveguide Bragg beam splitters, which is important for their efficient work. That is a general condition for all the waveguide interferometers considered in this paper. The bottom waveguide consists of two parallel vertical sides of height $H-L/2$ and the round bottom of radius $R=L/2$. To get equal propagation times of atoms along the two waveguides in presence of gravity, the total height of the bottom waveguide should be equal to $H=gT^2/8+L/2$, where $T=L/v_{wg}$ is the common propagation time. The main advantage of such a gravimeter over the Michelson waveguide gravimeter is that it does not need additional reflectors inside the waveguides, which produce additional phase shifts of the partial matter waves. For $L=1$ mm, $H=1.17$ cm, $v_{wg}=1$ cm/s, $F=10^5$ at/s and $T=0.1$ s the sensitivity of the gravimeter is equal to $4.81\times10^{-10}\:g/\sqrt{\text{Hz}}=0.48\:\mu{\text{Gal}}/\sqrt{\text{Hz}}$.

Another atomic interferometer, which greatly benefits from the use of atomo-photinic waveguides, is a Sagnac waveguide interferometer. One possible design of such a waveguide interferometer is shown in Fig. 3d. The interferometer consists of two waveguides. One of the waveguides has a shape of a closed loop, the ends of which cross each other at a right angle. This atomo-photonic waveguide is used for guiding of two-colour light modes and atoms along the loop. Unfortunately, the interference of guided light in the cross-region can't be used as a beam splitter of the Sagnac interferometer. The problem is that the corresponding optical lattice in the crossed region is oriented in such a way, that it does not provide the right splitting of an input atomic wave between the two ends of the loop. Instead, it splits part of the atomic wave into the second input port of the interferometer. To avoid that, the interference of optical modes of atomo-photonic waveguide in the cross region can be suppressed by using modes with TE polarization. To set up a right beam splitter for the Sagnac interferometer, an additional waveguide with a single-frequency standing optical wave can be used. This waveguide intersects the centre of the intersection region of the loop waveguide, such as its angles to the two input ports are $\pm$45 degrees. An evanescent wave optical lattice of this waveguide should provide splitting of atomic wave between the two ends of the loop of the Sagnac interferometer, such as one of the partial atomic waves propagate clockwise along the loop and another one anti-clockwise. Note, that the wavelength of light inside this waveguide can be different from the wavelength of light inside the atomo-photonic loop waveguide. Ideally, this wavelength should be short to reduce the period of the corresponding optical lattice and thus to increase the momentum of diffraction splitting the matter wave. In principle, the same type of beam splitter can be used for all waveguide interferometers considered above. A rotation of such an interferometer with angular rate $\Omega$ due to the Sagnac effect leads to the accumulated phase difference of $\Delta\phi=(1+3\pi/4)m \Omega L^2/\hbar$ between the two partial waves propagating along the loop in opposite directions, where $L$ is the size of the loop. The corresponding maximal sensitivity of the gyroscope to rotation for $L=1$ cm and $F=10^5$ at/s is  $S_{\Omega}=1.1\times10^{-8}$ rad/s/$\sqrt{\text{Hz}}$. The main advantage of such a Sagnac waveguide atom interferometer is that it is not sensitive to its linear accelerations. That is the main advantage of such an atomic gyroscope over gyroscopes based on Mach-Zehnder interferometers.

Let us now discuss a possible design of all-optical atom chip devices and describe the main challenges to their realisation. Fig. 4 shows a general structure of an integrated all-optical waveguide Mach-Zehnder interferometer. It includes two double-crossing curved waveguides, a surface source of ultra-cold atoms and integrated detectors of atoms at the ends of both waveguides. The possible design of waveguides was already discussed above, designs of a surface source of ultra-cold atoms and integrated atomic detectors need to be developed. To make this all-optical atom chip operational, it is necessary to populate one of the input atomo-photonic waveguides of the interferometer with ultra-cold atoms. The most obvious solution is to use a surface evanescent bichromatic light wave trap at one of the input waveguides, which would feed it with ultra-cold atoms. As cross-sections of the evanescent light wave atomic waveguides are rather small \cite{15}, such waveguides would benefit from high densities of ultra-cold atomic ensembles in the surface atomic source. One way to achieve that is to use atomic Bose-Einstein condensates (BEC) \cite{26}. The fastest and easiest way to produce an atomic BEC is an all-optical method of condensation \cite{27,28}, where thermal atoms are evaporatively cooled in an optical dipole trap. Such a BEC can be prepared in a free space next to the atom chip in a superposition of a standard magneto-optical trap (MOT) and an optical dipole trap. After that, the BEC can be transferred to the surface evanescent wave trap (SET) with a moving optical lattice \cite{29}. That process of transfer atoms into the SET can be extended in time to provide a quasi-continuous flux of atoms in the optical waveguides of corresponding integrated interferometers. The design of SET should provide also a continuous acceleration of atoms along the corresponding integrated waveguide to a certain velocity, which is needed to successful Bragg splitting at the beam splitters of the waveguide interferometer. This can be achieved with additional waveguides, which produces a moving evanescent wave optical lattice inside the SET.

\begin{figure}
	\centering
	\includegraphics[scale=0.35]{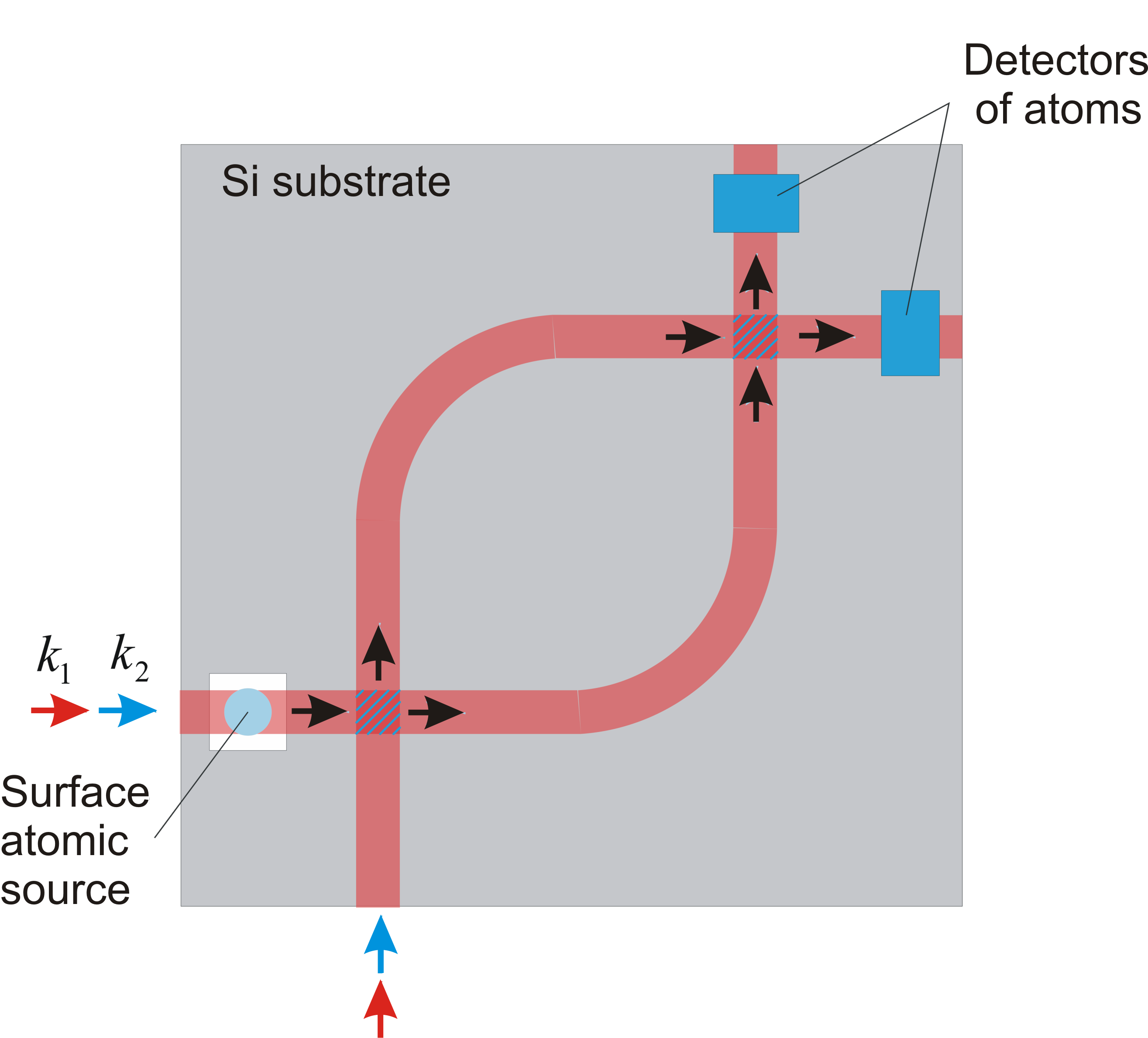}
	\label{}
	\caption{General design of an all-optical atom chip with a Mach-Zehnder waveguide atom interferometer.}
\end{figure}

Another important part of the all-optical atom chips are detectors, which have the ability to detect the relative numbers of atoms at the two output atomo-photonic waveguides of integrated interferometers. These detectors can be based on different physical principles, but the easiest solution is to use detectors based on the interaction of atoms with light. Note, that in a case of atomo-photonic waveguides for such detectors can be used evanescent light waves, which simplifies the whole design of the chips. Such detectors can be based on the absorption or fluorescence of the resonant detection light. One example of such detectors integrated into atom chips is based on integrated optical fibre cavities \cite{30}. Another promising approach to such detectors are plasmon-based waveguide sensors \cite{31}, which can be also non-destructive \cite{32}. Note, that the non-destructiveness of the detection process is very important in the case of using the integrated interferometers in a continuous regime, which is a key advantage of these types of atom interferometers.

In conclusion, designs and expected performances of different integrated atom interferometers based on atomo-photonic waveguides are described. It is shown that the corresponding inertial sensors based on the all-optical atom chips can have sensitivities to inertial forces, which are comparable to the sensitivities of the best quantum inertial sensors based on free-falling atoms. A possible design of the corresponding all-optical atom chip and its main elements are discussed.

\subsection{Acknowledgments}

We acknowledge the support of the UK government department for Business, Energy and Industrial Strategy through the UK national quantum technologies programme. Many thanks to Hugh Klein and Geoffrey Barwood for their comments and corrections to the paper.

\subsection{Data Availability}

The data that support the findings of this study are available from the corresponding author 
upon reasonable request.

\subsection{Author Declarations}

The authors have no conflicts to disclose.

\end{document}